\documentstyle[sprocl2,epsfig]{article}
\def\be{\begin{equation}}
\def\ee{\end{equation}}
\def\bea{\begin{eqnarray}}
\def\eea{\end{eqnarray}}

\begin{document}

\title{Spatial 't Hooft loop to cubic  order in hot QCD}

\author{P. Giovannangeli}
\address{Centre Physique Th\'eorique au CNRS,  
Case 907, Luminy 
F13288, Marseille, France\\
E-mail: giovanna@cpt.univ-mrs.fr} 

\author{C. P. Korthals Altes}

\address{CERN TH division,  
CH-1211 Gen\`eve 23, 
Switzerland\\
and\\
Centre Physique Th\'eorique au CNRS, Case 907, Luminy,F13288, Marseille, France\\ E-mail:altes@cpt.univ-mrs.fr}  


\maketitle

\abstracts{Spatial 't Hooft loops of strength $k$ measure the qualitative change in the behaviour of electric colour flux in confined and deconfined phase of SU(N) gauge theory. They show an area law in the deconfined phase, known analytically to two loop order with a ``k-scaling'' law $k(N-k)$. In this paper we compute the $O(g^3)$ correction to the tension. It is due to neutral gluon fields that get their mass through interaction with the wall. The simple k-scaling  is lost in cubic order. The generic problem of non-convexity  shows up in this order and the cure is provided. The result for large N is explicitely given. We show that nonperturbative effects appear at $O(g^5)$.}
\section{Introduction}

The deconfining phase of QCD describes  a plasma of in principle nearly free 
quarks and gluons. The colour electric flux of quarks and gluons is no longer
confined in tubes, but is Debye screened. The pressure is then approximately equal to that of a Stefan-Boltzmann gas, with the number of internal degrees of freedom  of gluons ($N$) and of
 quarks ($N_f$) appearing :

$$p_{SB}=(N^2-1+{7\over 8}N_f){\pi^2\over{45}}T^4$$. 

 Of course there are interactions, becoming stronger as we go down in temperature. As the critical temperature $T_c$ is on the order of $\Lambda_{\overline{MS}}\sim 200MeV$ the gauge coupling constant $g^2(T)$ is still $O(1)$ at $T\sim 4T_c$ which is where experiment may eventually take us. Unfortunately at these temperatures the perturbative pressure is wildly varying when adding higher order terms~\cite{arnold}~\cite{zhai}.
Since long one suspects the contributions from the long distance scales (Debye and magnetic screening lengths) to be the culprits for this failure of perturbation theory~\cite{recentkaj}. Lattice  evidence~\cite{hartowe}~\cite{hartlaine} suggests the series behaves well,  once the order where magnetic scales appear has been taken into account. A prime example is the spatial Wilson loop: its dominant contribution is from magnetic scales, and it is accurately described down to $T\sim 2T_c$~\cite{karschtension} by letting the coupling run due to the modes of order $T$. 

Another example is the Debye mass. To lowest order it is given by excitations on the scale $T$. But already in next to leading order the magnetic scales contribute~\cite{rebhan}. Once they are taken into account (and for any reasonable temperature they dominate~\cite{kajantie}) the remnant of the series is small according to numerical simulations~\cite{hartowe}~\cite{hartlaine}.  
   
The magnetic scales contribute a partial pressure of order $(g^2T)^3$, so come in only at $O(g^6)$ in the pressure~\cite{linde}. They give a non-perturbative contribution to this coefficient, which is under study~\cite{karsch}~\cite{recentkaj}.

So the data seem to tell us that the magnetic sector, in establishing a starting point for the series, is just as important as the electric sector with its gluonic quasi-particles.

A rationale may be provided by simply postulating magnetic quasi-particles~\cite{giovanna}  with a density $\sim (g^2T)^3$. A simple consequence is k-scaling for spatial Wilson loops at very high T. This was confirmed by lattice data~\cite{teper} to one percent accuracy.

In this context it is of interest to compute analytically the low orders of the  spatial 't Hooft loop, to compare the result to lattice simulations, so as to see from what order on the series behaves well.

 This loop measures the colour electric flux going through it.
One can write it as an electric dipole sheet in say an x-y plane at some fixed z and t coordinate. On the lattice the dipole sheet appears as 
 an electric twist on that same  x-y cross section of  the periodic box. 

The pressure is not changing, but a boundary effect occurs, proportional to the area of the twist. The effect of the electric dipole sheet will extend over a distance typically of the electric screening.  So it is reasonable to expect that  the magnetic scales, much larger then the Debye screening, will appear only through their effect on the Debye screening (which at $T\sim 2T_c$ is dominating the lowest order Debye screening~\cite{kajantie}). We find that this effect comes in at $O(g^5)$, see section(\ref{sec:semiclass}).

Hence we expect to produce a  purely perturbative series up and including $O(g^4)$. Numerical simulation is available~\cite{deforcrand} from very near $T_c$, where perturbative methods are certain to fail, to about $2.5 T_c$, where the data are in reasonable agreement with two loop results~\cite{bhatta}. In order to do a more meaningful comparison lattice simulations of the loop from $2.5T_c$ on should become available.  

Also in this paper we extend the computation of the 't Hooft loop
from the known two loop order $g^2$ to include all $g^3$ effects. 
Apart from the motivation above, this computation serves a two-fold purpose. First it will verify or falsify  
the simple k-scaling in the strength of the loop, found to one and two loop order. 

Second there is a general motivation. The computation of the tension
involves a potential between two degenerate minima, hence a potential with a non-convex part. If the strength of the loop is $k$ the profile induces masses
 of the order $O(1)$ in the coupling in $2k(N-k)$ gluon fields. The others stay massless, since they do not interact with the wall to one and two loop order, so do not contribute to the tension.
This is the cause of ``k-scaling''.
 But in three loop order they can develop a selfenergy, through which they start to interact with the wall.  The order $g^3$ involves the the resummation of the Coulomb propagator by this selfenergy. The latter
is related to the second derivative of the potential, hence with a negative part. The strength of the loop enters in this selfenergy and it turns out that
to leading order in the number of colours there is a window in the strength ${k\over N}$ where the selfenergy stays non-negative and where we can solve for the tension. That is what we will do in this paper. 

On the other hand the generic solution to the problem is of course suggested by the physics of the problem: the selfenergy is a consequence of the collective interaction of the wall with the massless gluonfield. Hence we have to substitute the profile into the selfenergy and solve for the eigenvalue spectrum of the  Schroedinger equation. It turns out this spectrum can be found with the methods of reference~\cite{dashen}. Hence it is interesting to see how the two approaches compare in their common domain of validity. This will be done in a sequel to this paper~\cite{korthalsgiovanna}.

The lay-out of the paper is as follows. Section(2)
sets the problem in the familiar context of a spontaneously broken centergroup symmetry. Then in the next section the definition of the 't Hooft loop is given, and an intuitive argument for its behaviour in hadron and plasma phase.
 
Then we start the main thrust of the paper in section(\ref{sec:semiclass})
with an outline of our semiclassical approach and the necessary tools.
One, and two loop results are reviewed in the next sections. The infrared divergencies in the three loop case are  discussed and how to get rid of them respecting centergroup invariance. Then in section(\ref{sec:g3}) we compute the 
$O(g^3)$ terms for large N and a window of 't Hooft loop strengths.

In section(\ref{sec:discussion}) the results are discussed. Appendices 
contain technical details.

\section{Z(N) symmetry in hot gluodynamics}\label{sec:symm}

 In absence of matter fields transforming non-trivially under the 
centergroup $Z(N)$ of $SU(N)$ there are two distinct canonical symmetries:
magnetic and electric Z(N) symmetry~\cite{thooft}~\cite{korthalskovner}. They act through charge operators on the quantum states of the theory.

There is another $Z(N)$ symmetry~\cite{thooft} induced by a compact periodic dimension. Only the timelike  case is here  of interest. It is a symmetry of the pathintegral giving  the pressure of the theory. This  path integral is periodic in Euclidean time. It acts on the fields of the 
path integral as a gauge transformation $\Omega_l$ with a discontinuity $\exp{il{2\pi\over N}}$  in the centergroup. Since the fields do by assumption not
feel
the action of the centergroup their periodicity is respected by this gauge transformation. The action is invariant and so is the integral. To make this into
something more useful we have to have an order parameter, that does transform non-trivially.

This is the timelike Polyakov loop:
\be
P(A_0)={1\over N}Tr{\cal{P}}\exp{i\int d\tau A_0(\tau,\vec x)}
\ee
It transforms in an obvious way:
\be
P(A_0^{\Omega_l})=\exp{il{2\pi\over N}}P(A_0)
\ee
 So the constrained path integral we get by integrating over all configurations
with a fixed value of the Polyakov loop will then have the same value in all
Z(N) transformed points, because only the loop is transforming. Define the 
effective action, with matter fields represented by dots:
\be
\exp{-{1\over {g^2}}U(\widetilde P)}=\int DA ...\delta(\tilde P-P(A_0))\exp{-{1\over {g^2}}S(A,...)}
\ee
then 
\be
U(\widetilde P)=U(\exp{il{2\pi\over N}}\widetilde P)
\ee
 
The effective action is $Z(N)$ invariant.
 $\widetilde P$ is the VEV of the Polyakov loop. If the VEV is non zero, like
in the deconfined high T phase, we have a Z(N) periodic potential for the VEV.

This potential is quite useful for the physics of the plasma. First of all it can be seen on general grounds, that the potential must have a minimum for the space averaged loop $\widetilde P=1$~\cite{gocksch}. The potential is known to two loop, $O(g^2)$, and we will compute it below to $O(g^3)$. It is gauge invariant by construction and may serve as a {\it{definition}} of the free energy at $\widetilde P=1$. This definition has the advantage that the Feynman rules have a built-in gauge invariant infrared cut-off. It is given by by the scale $qT$, which determines the deviation of $\widetilde P$ from its bulkvalue 1: $\widetilde P\sim\exp{-qT/T}$.

\section{The 't Hooft loop and colour electric flux}\label{sec:loop}

The behaviour of colour electric flux in the confined and deconfined phase is
expected to be qualitatively very different. In the confined phase we expect
  rings of electric fluxtubes to be the glueballs, whereas in the deconfined phase
the gluons are the electric quasi-particles: approximately a free Bose gas.

The 't Hooft loop $V_k(L)$, the closed loop $C$ say being square, size  LxL in the x-y plane, is usually defined~\cite{thooft} as a
closed magnetic flux loop $L$ of strength $k{2\pi\over N}$, with N the number of colours. In operator language it is defined as a gauge transformation with
a discontinuity   $\exp{ik{2\pi\over N}}$ when crossing the minimal surface.
As we will justify below, one can write this gauge transform as a dipole sheet~\cite{korthalskovner}~\cite{kovnerrosen}, see fig.(\ref{fig:dipole}):

\begin{figure}
\begin{center}
\includegraphics{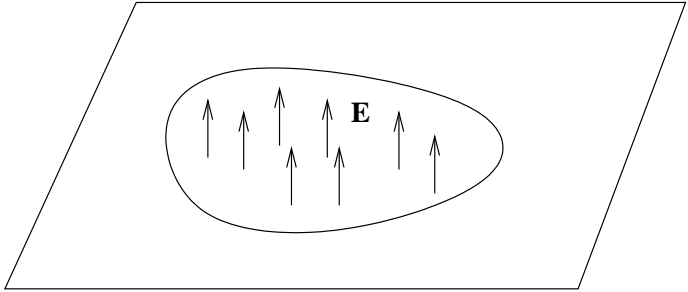}
\caption{The 't Hooft loop in an x-y cross-section of the box. Arrows indicate the electric fieldstrength projected on the hypercharge $Y_k$. Eventually we will consider the loop stretched to the full cross-section.}
\label{fig:dipole}
\end{center}
\end{figure}

\be
V_k(L)=\exp{i{4\pi\over N}\int_{S(L)} dxdyTrE_zY_k}
\label{eq:loop}
\ee

where the $NxN$ diagonal traceless matrix $Y_k$ is defined as
\be
Y_k=diag (k,k...k,k-N,k-N.....,k-N)
\label{eq:charge}
\ee
with N-k entries k and k entries k-N, to have a traceless matrix.
The charges $Y_k$ are generalizations of the familiar hypercharge, with k=1.
The charge $Y_k$ of a gluon is 0 or $\pm N$. The multiplicity of the value $N$ is $k(N-k)$.  The same is true for the value $-N$.  So, e.g.  for $N=3$ and $k=1$ one finds the four kaons with hypercharge $\pm 3$.

Exponentiation of  $Y_k$ gives $\exp{i{2\pi\over N}Y_k}=\exp{ik{2\pi\over N}}\equiv z_k$, the centergroup element.
 
$E_z$ is the z component of the canonical electric field strength operator
$\vec E=\lambda_a \vec E$ \footnote{ The $\lambda$ matrices being normalized to 
$Tr\lambda_a\lambda_b={1\over 2}\delta_{a,b}$.}. 

The 't Hooft loop $V_k(L)$ can be written as  $\exp{i{2\pi\over N}\Phi_k}$ with
\be
\Phi_k= \int_{S(L)} dxdy2TrE_zY_k
\label{eq:eflux}
\ee

\noindent the electric flux operator. 

If a spatial Wilsonloop in the fundamental representation - a fundamental electric flux loop - $W(L')=\exp{i\int_{L'}d\vec l.\vec A}$ is piercing through the area subtended by the 't Hooft loop, the loops
do not commute anymore with each other.  The effect 
\footnote{ through the canonical commutator $[E^a_l(\vec x),A^b_n(\vec y)]=i\delta^{a,b}\delta_{l,n}\delta(\vec x-\vec y)$.} is a phase $\exp{ik{2\pi\over N}}$ and can be
written as the 't Hooft commutation relation:
\be
V_k(L)W(L')V_k(L)^{-1}=\exp{i{2\pi\over N}Y_k}W(L')
\label{eq:thcommutation}
\ee

The appearance of the phase in the commutation relation shows that the 't Hooft loop is a gauge transform with a discontinuity $\exp{i{2\pi\over N}}$, detected by the Wilsonloop. 

The choice of the charge $Y_k$ is determined up to a regular gauge transformation. That will not change the commutation relation above. In particular a permutation of the 
diagonal elements will leave $z_k$ invariant. The reader might object
that to a given centergroup element there corresponds a lattice of points in Cartan space. So which point to choose?
The answer is: the assignment of a specific value of $z_k$ to the plasma groundstate
is arbitrary. But the jump at the dipole layer must be such that one arrives at
a nearest neighbour of the chosen lattice point~\footnote{An intriguing explanation has been given by Kovner~\cite{kovnerfossil}.}.

\subsection{Thermal expectation value of the 't Hooft loop in the quasiparticle picture}\label{subsec:quasi}

Let us imagine the hadronphase at some temperature T - below $T_c$ - as a gas of
microscopic closed fundamental flux loops, i.e.  Wilsonloops on the scale of glueballs. Then, from the commutation relation above, we see that 
only  glueballs sitting on  the perimeter $L$ of the 't Hooft loop will have an effect on the 't Hooft loop.

But, on the contrary, a gas of free gluons will change the behaviour of the 
't Hooft loop into an area law behaviour. With the tension $\rho_k(T)$ and the area $A(L)$:
\be
<V_k(L)>=\exp{-\rho_k(T)A(L)}
\ee

This area law follows from the quasi-particle picture in the following simple way.

 Just one single gluon will change the value of the loop by a minus sign.
To see this, note that the gluon has a charge 0 or N. This is true for any value of k in view of the definition
of k charge, eq.(\ref{eq:charge}). What changes is the multiplicity of the charge $\pm N$. We know this charge is screened by the plasma, with a screening length $l_E$. So if the particle is within distance $l_E$  from the surface spanned by the loop one half of the flux will go through the  loop, the other 
half does not. As a result the loop captures a flux, eq.(\ref{eq:eflux}):
\be
\Phi_k={N\over 2}
\ee
and the loop $V_k(L)=\exp{i{2\pi\over N}\Phi_k}$ acquires the value $-1$.

Let $P(l)$ be the probability that $l$ gluons are in the slab of thickness $l_E$. We suppose this probability is centered around the average $\langle l\rangle$, with width
proportional to $\langle l\rangle$, as behooves a thermodynamic distribution for a gas of 
bosons with a mass. The outcome  for the average is then:
\be
\sum_l(-)^lP(l)=\exp{-c<l>}
\ee
with the proportionality factor $c$ related to the shape of the distribution~\footnote{The distribution for massless bosons does not have this shape and does not give an arealaw.}.
This result is true for one gluon species with charge $\pm N$.
We know from eq.(\ref{eq:charge}) there are $2k(N-k)$ of such species. Since they are acting independently
 the tension becomes:
\be
\rho_k=c2k(N-k)l_E n(T)
\ee

The dependence on the strength k is the k-scaling law for 't Hooft loop
 tension. The screening length is proportional to ${1\over{\sqrt{g^2N}T}}$.

Summing up: a gas of free gluons predicts the ratio $\rho_k/\rho_1=k(N-k)/(N-1)$. The factor $k(N-k)$ is just the number of gluons that have  $Y_k$
charge $N$.

\section{Semi-classical determination of the 't Hooft loop expectation value}\label{sec:semiclass}

The arguments in the previous section were semi-classical, in that only the
density $n(T)$ of the free gluons is quantum mechanical. So we can expect that a  semiclassical
approach would be in place with a systematic expansion in the coupling. 

Apart from its obvious connection with colour electric flux the 't Hooft loop
is intimately related to the Z(N) symmetry of section(\ref{sec:symm}). Take its order
parameter $P(A_0)$, and move it through the dipole layer. Then it will get
multiplied by the discontinuity $\exp{ik{2\pi\over N}}$, being a (heavy) fundamental test charge.
 Immersing the loop in the plasma induces a disturbance. The disturbance is described by a profile $C$. Since the loop is gauge invariant the response of the plasma is too. This profile is 
the phase of the Polyakov loop as a function of its distance to the minimal surface.

So the approach will be to compute the free energy excess $\Delta F$ due to the presence
of the Polyakov loop profile. Let the box be of size $L_{tr}^2xL_z$, with $L_z>>L_{tr}$
and both macroscopic.
Extend the loop in fig.(\ref{fig:dipole}) to the full x-y cross-section of size $L_{tr}^2$ and located at say $z=0$.

Then we have $\exp{-\Delta F(C)}=\int DC\exp{-{L_{tr}^2\over{g^2}} U(C)}$. $U(C)$ is the constrained effective potential we mentioned in section(\ref{sec:symm}).

We will minimize this excess free energy by varying the profile $C$, find the profile $C_{min}$ and
the minimum free energy $U(C_{min})$, which is the tension of the 't Hooft loop. 

This is a simple procedure. What is technically involved is the transformation
of the gauge potential to the gauge invariant Polyakov loop. This is done in the next subsection.

\subsection{The constrained pathintegral}\label{subsec:constrained}

So the thermal average of our loop is given by the Gibbs sum
\be
<V_k(L)>=TrV_k(L)\exp{-H/T}
\ee

The average can be transformed into Euclidean path integral form.
When doing so, one has to realize that the operator 
$\exp{i{4\pi\over N}\int_S(L) dxdyTrE_zY_k}$ is only gauge invariant 
on the physical subspace. This follows from it being a gauge transform
with a discontinuity in the center of SU(N). The discontinuity commutes with
all of the gauge group.  If we chop up the operator  in 
little Euclidean time bits $\delta \tau$, then every bit apart will not be a centergroup element, hence will not be gauge invariant!. This in contrast to the Boltzmann factor
where this operation will not affect the gauge invariance. So the operator stays inserted at a fixed time slice, say $\tau=0$. Only there $A_0\neq 0$, for all other times $A_0=0$. We get:
\be
<V_k(L)>=\int DA_0 D\vec A\exp{-{1\over{2g^2}}\int d\tau d\vec x\big(Tr(\sum_i(F_{0i}-\delta(\tau)s_i^k)^2+{\vec B}^2\big)}
\label{eq:pathint}
\ee

The terms in the exponential constitute the action $S(A,s_k)$ with the original surface distribution of dipoles as a source $s^k_i\equiv 2\pi Y_k\partial_i \theta(z)$ localized at $z=0$.  This source induces a profile in the thermal Polyakov loop, or more precisely in the average over transverse directions  $\bar P(A_0)\equiv{1\over{L^2}}\int dx dy P(A_0)$. 
 
The source term tells the Polyakov loop to jump over $P({2\pi\over N}Y_k)=\exp{ik{2\pi\over N}}$ at the minimal surface and to have value $1$ at long (longer than the Debye length $l_E$ distance from the 
surface. 

So the thermal average becomes in terms of the profile $\widetilde P(z)$:
\be
<V_k(L)>=\int D\widetilde P(z)\exp{-{L_{tr}^2\over{g^2}}U\big(\widetilde P\big)}
\label{eq:profileint}
\ee
The effective action $U\big(\widetilde P\big)$ is given by the constrained 
path integral we introduced in section(\ref{sec:symm}):
\be
\exp{-{{L_{tr}^2\over{g^2}}U\big(\widetilde P\big)}\equiv \int DA_0 D\vec A\delta\big(\widetilde P-\bar P(A_0)\big)\exp{{1\over{g^2}}S(A,s_k)}}
\label{eq:constraint}
\ee
and we have used an abbreviated notation for the constraint:
\be
\delta\big(\widetilde P-\bar P(A_0)\big)=\Pi_{z,l}\delta\big(\widetilde P^{(l)}(z)-\bar P^l(A_0(z))\big)
\ee
where $l$ runs from 1 to $N-1$.
In the following we parametrize the fixed loop by a diagonal traceless $NxN$ matrix $C(z)=diag(C_1,C_2,.....C_N)$ :
\be
\widetilde P^{(l)}\equiv {1\over N}Tr\exp{ilC(z)}
\ee
So the effective potential is defined on the Cartan space in which the matrices $C$ live. The matrix describes the profile of the loop.
Once again, the way the strength $k$ comes into the effective action is through
the boundary condition at the dipole layer.

\subsection{Thermodynamic limit and minimizing path}\label{subsec:thermolimit}

In the thermodynamic limit $L_z$ and the transverse size $L_{tr}$ become very large with respect to any microscopic scale. 

Then the integral over profiles, eq (\ref{eq:profileint}), gets its main contribution from the minimizing configuration $C_{min}$. Writing the fluctuations as 
$C(z)=C_{min}(z)+\gamma(z)$ we obtain:
\be
<V_k(L)>=exp{-{L_{tr}^2\over{g^2}}U(C_{min})}\int D\gamma(z)\exp{-{L_{tr}^2\over{g^2}}\gamma(z).U''(C_{min}).\gamma(z') +O((\gamma)^3)}
\ee

So the tension will become:
\be
\rho_k=U(C_{min})+{1\over{L_{tr}^2}}Tr log U''(C_{min}))
\label{eq:correction}
\ee
The stability matrix $U''$ has non-negative eigenvalues for all fluctuations in the 
profile, because our kink $C_{min}$ is stable~\footnote{No zeromode appears
because  the profile location is fixed by the fixed loop.}.

The strength of the loop determines the jump at the dipole layer. It also fixes the 
path in Cartan space (the space of diagonal $C$'s) along which the minimal
profile is realized. 
The path turns out to be the  simplest possible~\cite{giovanna}:
if parametrized by $q$, $0<q<1$, it is given by the one dimensional set
of Cartan matrices 
\be
Y_k(q)=qY_k 
\label{eq:path}
\ee
\noindent with $Y_k$ the charge characterizing the 
strength of the dipole layer, eq.(\ref{eq:loop}).
 In exponentiated form it goes from 1 to
$\exp{ik{2\pi\over N}}$, as $q$ goes from $0$ to $1$.
Z(N) invariance of the profile functional $U(C)$ garantuees we we can
take a smooth path from 1 to $\exp{ik{2\pi\over N}}$, instead of a path
that makes a jump $\exp{ik{2\pi\over N}}$ at the loop and returns to 1.
A  proof of the rectilinear path being the minimal one is still lacking.
Only for SU(3) and SU(4) it is known to be the case by inspection~\cite{bhatta}~\cite{giovanna} and a proof at large N is given in ref.~\cite{bhatta}.
 
\subsection{The gradient expansion of the effective action}\label{subsec:grad}

In principle the tension is given by eq.(\ref{eq:correction}). We have now
to work out an approximation scheme in weak coupling.

The effective action appears in the constrained path integral. This path integral contains the profile $\widetilde P(z)$ as parameter. As long as we limit ourselves to profiles with gradients much smaller than the profile, a gradient expansion in $\partial_z\widetilde P(z)$ should make sense.

To see this go from the $A_0=0$ gauge in eq.(\ref{eq:pathint}) to background gauge.
So the potentials and therefore the  Polyakov loop are parametrized through a colour diagonal background field:
\be
A_{\mu}=B(z)\delta_{\mu,0}+gQ_{\mu}(\tau,\vec x)
\label{eq:quantumfluc}
\ee
and choose background gaugefixing  $Tr(D(B)_{\mu}Q_{\mu})^2$.
A systematic loop expansion of the constrained path integral is possible~\cite{korthalsaltes}.

To lowest order the background field $B$ equals the phase $C$ of the  Polyakov loop profile through 
the constraint in the pathintegral, eq.(\ref{eq:constraint}). The phase $C$
appears in all derivatives $D_0(C)$. Since the phase is diagonal in colour 
the colour basis for the quantum fluctuations in eq.(\ref{eq:quantumfluc})
that diagonalizes the covariant derivative is the Cartan basis: $\lambda^{ij}$ are the N(N-1) matrices with only one entry $(i\neq j)$ non-zero and equal to ${1\over{\sqrt 2}}$. Then the (N-1) diagonal matrices $\lambda^d$ with the first d-1 entries equal to 1 , the d'th entry equal to 1-d, and all others equal to 0. The normalization stays ${1\over 2}$ as for the Gell-Mann matrices. The fluctuating fields $Q$ can be decomposed on this basis:
\be
Q=\sum_{ij}Q^{ij}\lambda_{ij}+\sum_dQ^d\lambda_d
\ee

 The covariant derivative will act as $D_0(C)=\partial_0+i(C_i-C_j)$ on the $Q^{ij}$ components and like $D_0(C)=\partial_0$ on the diagonal components. So the Feynman rules are the usual
background field rules, with the Matsubara frequency $p_0$ shifted by the phase
 of the Polyakov loop. The phase $C(z)$ commutes with $\partial_z$ in the gradient expansion. For the Coulomb part of the propagator we use the resummed propagator: $({\Delta^{ij}_{00}})^{-1}=\vec p^2+m_D^2+T^2(C_i-C_j)^2$ for the $Q^{ij}$ propagators and $({\Delta^{d}_{00}})^{-1}=\vec p^2+m_D^2$  for the $Q^d$ propagators.

We are interested in the effective potential along the minimizing path $Y_k(q)$. Then a  simplification occurs in the Feynman rules. Along that path we have
$D_0(C)=\partial_0\pm i2\pi q$ for those gluon fields $Q_{ij}$ with charge $\pm N$. If the charge is zero the covariant derivative reduces to $\partial_0$. 
So in fig. (\ref{fig:12loop}),(\ref{fig:Df}), (\ref{fig:Dp}) and (\ref{fig:self}) the lines and vertices obey the above rules.

\subsection{One loop tension}\label{subsec:oneloop}
  
The only classical term that survives in the effective action is the electric field strength:
\be
U(C)=U_{cl}(C)={1\over{g^2}}Tr(\partial_z C)^2
\label{eq:classical}
\ee
No potential term survives in $U_{cl}(C)$. And this means there is no 
area law in this classical approximation:
in our box of size $L_zL_{tr}^2$ the profile minimizing $U_{cl}$ is linear in
$z$. Hence the profile yields a flat action density, so no wall results. 

\begin{figure}
\begin{center}
\includegraphics{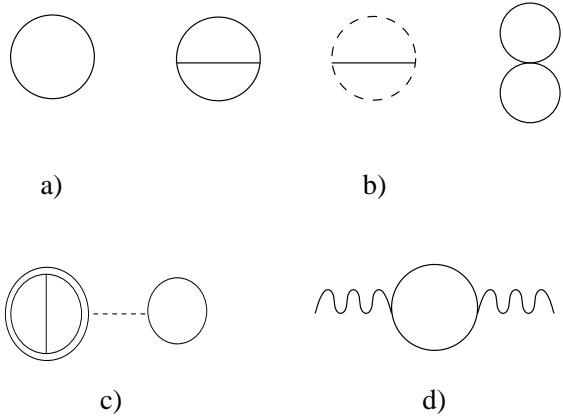}
\caption{(a) is the one loop contribution, and stands for gluon and ghost loop. (b) are the two loop contributions.
(c) is the renormalization of the Polyakov loop (double circle) inserted  (dotted line) into the one loop, see text below eq.(\ref{eq:twolooppot}).
(d) is the kinetic term with the loop being a gluon or ghost loop}
\label{fig:12loop}
\end{center}
\end{figure}

So we have to expand to one loop order from the graph in fig.(\ref{fig:12loop}a). 

The gradient expansion gives a
term proportional to $Tr(\partial_z C)^2$, and a potential term without
gradients:

\be
U(C)=\int dz \big({1\over{g^2}}K(C)Tr(\partial_z C)^2+V(C)\big)
\label{eq:1loop}
\ee
To find the minimum configuration for this form of action is well known~\cite{bhatta}. The result is:
\be
{1\over g}\partial_zC(K(C))^{{1\over 2}}-(V(C))^{{1\over 2}}=0
\label{eq:eqnofmotiongen}
\ee
as equation of motion for $C=C_{min}$. 

The tension reads:
\be
\rho_k={2\over g}\int dC(K(C)V(C))^{{1\over 2}}
\label{eq:tensiongen}
\ee

We now write 
\bea
K&=&1+g^2K_1+\ldots{..}\\
V&=&V_1+g^2V_2+\ldots{..}
\eea
In the one loop approximation the potential term equals:
\be
V_1(C)={2\pi^2T^4\over 3}\sum_{ij}({C_{ij}\over{2\pi }})^2\big(1-|{C_{ij}\over{2\pi }}|\big)^2
\label{eq:onelooppot}
\ee

\noindent where $C_{ij}=C_i-C_j$.
 
Now we focus on the minimizing path $Y_k(q)$. Then, from the explicit form
of the charge $Y_k$, eq.(\ref{eq:charge}), we see that all non-zero
$C_{ij}=\pm 2\pi q$. As the effective potential $V_1(C)$ is even in $C_{ij}$ all terms
contribute equally.
 And there are $2k(N-k)$ pairs $ij$ with non-zero charge.
Similar counting applies to the classical kinetic term, and after these manipulations we have for the sum of the classical kinetic term and the one loop potential $V_1(C)$ on the path $Y_k(q)$:
\be
{T^2\over{g^2}}Tr(\partial_z C)^2+V_1(C)= 2k(N-k)\int dz\big({1\over{2g^2N}}(2\pi T)^2Tr(\partial_z q)^2+{2\pi^2T^4\over 3}q^2(1-|q|)^2\big)
\label{eq:lowestorder}
\ee

So we have found once more the k-scaling of the  quasi-particle picture in section(2)!
The quasi-particles with charge $\pm N$ correspond precisely to the fields with
massive propagators, the ``neutral'' particles to the massless propagators. Only massive propagators contribute to this order.

Searching for a minimizing profile $q_{min}(z)$
forces us to to choose a configuration
$q_{min}(z)$ with $\partial_z q_{min}(z)=O(g)$. This in order for the electric term to be of the same  order as the  one-loop term. So it vindicates the use of the gradient expansion. The dependence on $z$ of the profile is through the combination $gz$. Of course this was to be expected, because the electric dipole layer  disturbs the plasma only over the Debye screening distance.

Still we did not mention the boundary condition at the surface spanned by the loop.
As in a surface layer of dipoles in electrostatic Maxwell theory we expect the electric field to be continuous when going through the layer. This then fixes the minimal profile.

The equations of motion for the profile become
\be
\partial_zq -m_Dq(1-q)=0
\label{eq:eqmotion}
\ee
for $q=q_{min}$  with $m^2_D={g^2N\over 3}T^2$ the Debye mass. 

Note that the equations of motion and hence the profile are independent of the strength $k$.

The tension of the wall in this leading order is given by~\cite{bhatta}~\cite{giovanna} 
\be
\rho^{(1)}_k=k(N-k){4\pi^2\over{3(3g^2N)^{1\over 2}}}T^2
\label{eq:leadingtension}
\ee

The tension  has the same parametric dependence on N, k and the 't Hooft coupling $g^2N$ as we found in the quasiparticle picture. 

\subsection{Next to leading order contribution to the tension}

The correction to the leading result comes  from combining the one loop kinetic term with the two loop potential term evaluated in the background $q(z)$. This will give an $O(g^2)$ correction to the result above and has been discussed elsewhere~\cite{bhatta}~\cite{korthalsaltes}~\cite{giovanna}. 
Here we only look in some detail to the infrared aspects. 

To begin, the two loop correction $V_2(q)$ to the potential is 
infrared finite and equals (see fig.(2b)) 
\be
g^2V_2(q)=-5{g
^2\over{(4\pi)^2}}k(N-k)q^2(1-|q|)^2 
\label{eq:twolooppot}
\ee
Important is to realize that the Feynman rules from the constrained effective action 
 eq.(\ref{eq:constraint}) are not only involving the traditional vertices, but extra vertices due to the contraint $\bar P-\bar P(A_0)$. They come on one hand from expanding $\bar P(A_0)$ in the quantum field, and on the other hand from insertions into traditional vertices.  They produce the effects of the renormalization of the Polyakov loop. In the two loop
approximation they give rise to only one diagram, representing the effect of
the one loop renormalization of the Polyakov loop inserted in the one loop free energy, fig.(\ref{fig:12loop}c). 

The result from fig.(\ref{fig:12loop}c) was discussed in ref.~\cite{bhatta}and in a systematic way in ref.\cite{korthalsaltes}.
As far as the tension concerns there is no qualitative difference with the leading term, and the result turns out to be again k-scaling~\cite{bhatta}~\cite{giovanna}(see  appendix A):
\be
\rho_k(T)=\rho^{(1)}_k(T)(1-{\alpha_s\over{4\pi}}N(15.2785...-{11\over 3}(\gamma_E+{1\over{22}}))+O(g^3)
\label{eq:tensionordertwo}
\ee
The coupling is now running, due to the one loop kinetic term. It is defined in the $\overline {MS}$ scheme~\cite{recentkaj}.

\subsection{Profile to two loop order and Debye mass}

 The behaviour of the corrected profile far from the loop is less straightforward. This is obvious since the Debye mass controls the
behaviour there, and the next to leading correction to the Debye mass gets contributions from the electrostatic sector, and from  an infinite number of 
diagrams in the magnetostatic sector\cite{linde}. 
 
So let us look at the contribution from the static sector to the kinetic term $K_1(q)$ due to the diagram
in fig.(\ref{fig:12loop}d). From the Feynman rules in Feynman background gauge we find the   
only contribution  from a Coulomb and a spatial propagator, carrying mass $m_D^2+(2\pi Tq)^2$, $(2\pi Tq)^2$ respectively:

\be
g^2{\vec p}^2K_1(q)=4g^2N{\vec p}^2 T\int {d\vec l\over{(2\pi)^3}}{1\over{(\vec l^2+m_D^2+(2\pi T q)^2)((\vec l+\vec p)^2+(2\pi T q)^2)^2}}
\ee
The external momentum $p\equiv |\vec p|$ is $O(gT)$ as we argued above for the gradient of the profile. In this section the mass parameter $q=O(g^2)$ will be parametrically smaller than the Debye mass, hence we will neglect it in the Coulomb propagator. 
After integration one finds:
\be
{\vec p}^2K_1(q)=-4{\vec p}^2{g^2N T\over{8\pi}}\big({1\over{ip}}{ log{-ip+m_D+2\pi qT\over{ip+m_D+2\pi qT}}}\big)
\label{eq:cut}
\ee

\noindent where we put the cuts along the imaginary axis in the complex p-plane, starting from $\pm i(m_D+2\pi qT)$. 

So the pole $ip=m_D$ is well separated from the cut in eq.(\ref{eq:cut}), so we substitute $m_D$ for $ip$. 
 The gradient expansion for the static modes has changed the $O(g^2)$ graph into a term linear in the coupling $g$, multiplied by the logarithm. Put into the  equation of motion for the profile, eq.(\ref{eq:eqnofmotiongen}), it gives a correction to the Debye mass appearing in the lowest order equation of motion, eq.(\ref{eq:eqmotion}).  When $q=O(g^2)$ we get Rebhan's correction~\cite{rebhan} to the Debye mass: 
\be
m_D=\sqrt{{N\over 3}}gT+{g^2N\over{4\pi}}log({1\over g})
\label{eq:debyecorr}
\ee

There is also a static contribution in fig.(\ref{fig:12loop}d) from the spatial propagators only.
It reads
\be
g^2{\vec p}^2K_1(q)=-{1\over 3}g^2N{\vec p}^2 T\int {d\vec l\over{(2\pi)^3}}{1\over{(\vec l^2+(2\pi T q)^2)((\vec l+\vec p)^2+(2\pi T q)^2)^2}}
\ee
Doing the integral will give us an amplitude with cuts starting at $\pm i4\pi qT $
and our procedure will fail for $q$ on the order of the magnetic mass because the pole is now on the cut. This failure indicates that other means than perturbation theory are necessary~\cite{linde}. 

It should be emphasized that only when $q=O(g^2)$ we get this non-perturbative behaviour~\cite{linde}. For q parametrically
larger the result is still perturbative. So in the q-integration leading to the tension, eq.(\ref{eq:tensiongen}):
\be
\rho_k(T)={2\over g}\int dq (K(q)V(q))^{1\over 2}\sim{2\over g}\big(1+.....+
\int^{g^2}dq q g(log(1/g)+non pert)\big)                         
\ee
\noindent the non-perturbative contribution will  contribute to $O(g^5)$.

In conclusion, the next to leading order has electrostatic and magnetostatic divergencies that show up only in the profile at long distance, $m_D|z|=\log({1\over{g^2}})$, and in the tension at $O(g^5)$. They typically show up 
for profiles on the order of $q=O(g^2)$. At values of $q=O(1)$ the theory is protected from the infrared, to this order.  
This concludes the discussion of the next to leading order diagrams. 

\section{Linear divergencies in three loops and Z(N) invariant selfenergy}\label{sec:threeloop}

In this section we analyze the two loop kinetic term and the three loop
potential term. We look for infrared divergencies that might contribute 
to cubic order to the tension. It will turn out that none of those are present in the 
kinetic term, but indeed in the potential term. 

\subsection{Three loop potential and self energy matrix}\label{subsec:3looppot}

From the preceding section one would conclude that only the fringes of the wall, near the unperturbed plasma, have infrared divergencies. 
The  three loop potential, surprisingly, shows them  for {\it{all}} values of the profile~\footnote{In a previous paper~\cite{giovanna} we overlooked this possibility}. Let us look in more detail to this.

The three loop diagrams are divided in two sets, $D_f$ and $D_p$.
The diagrams in $D_f$ are all diagrams with the topology of free energy diagrams, but put in the background $q$ of the tunneling path. Their calculation involves only vertices from the Yang-Mills action, background gauge condition and ghost action. $D_p$  involves the vertices due to the constraint and renormalizes the Polyakov loop.

\begin{figure}
\begin{center}
\includegraphics{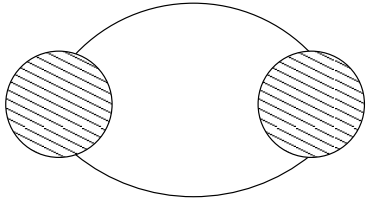}
\caption{The only  three loop diagram of  free energy topology with an infrared divergence. The shaded blob is the one loop selfenergy of fig.(\ref{fig:self}).The colour index of the two propagators need not be the same due to background dependence inside the blobs.}
\label{fig:Df}
\end{center}
\end{figure}

 The set $D_f$ is  divided in turn
into two-particle irreducible diagrams $D^{(2)}_f$ and one-particle irreducible diagrams $D^{(1)}_f$.  The two particle irreducible diagrams have been analyzed for $q=0$ and are individually infrared convergent. This means they are 
$O(q)$ for small $q$, like their two loop counterparts. There is a single  one-particle irreducible diagram as shown in fig.(\ref{fig:Df}).

\begin{figure}
\begin{center}
\includegraphics{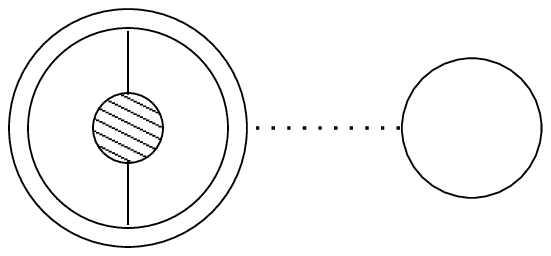}
\caption{Only three loop insertion diagram with infrared divergence}
\label{fig:Dp}
\end{center}
\end{figure}

This diagram is in the case of vanishing background linearly divergent for the
$00$ (``Coulombic'') components of the propagator. Similarly, the diagram in fig.(\ref{fig:Dp}) showing the self energy insertion on the one loop renormalization of the Polyakov loop has a linear divergence, which in the case of vanishing background has been analyzed by Gava et al.~\cite{gava}.

For notational convenience we write $\Pi_{c,c'}^{(k)}(q)$ for the Coulombic selfenergy matrix at zero momentum and frequency, dropping the Lorentz indices. The index k means we are evaluating the selfenergy along the tunneling path $Y_k(q)$.
If the two propagators  have no background
induced mass 
the diagram is linearly divergent in the infrared and we have to resum the propagator by including the selfenergy $\Pi_{c,c'}^{(k)}(q)$  at zero Matsubara frequency and momentum. At $q=0$ we have the Debye mass $\Pi_{c,c'}^{(k)}(0)=m_D^2\delta_{c,c'}$. So the resummation we used up to now subtracts correctly the selfenergy at $q=0$. 
But to have a finite result for $q\neq 0$ we have to subtract the selfenergy for all $q$, and hence resum it for all $q$. Only then the linear divergence is cancelled. So the inverse Coulomb propagator in the background $q$ becomes:
\be
\vec l^2\delta_{c,c'}+\Pi_{c,c'}^{(k)}(q)
\label{eq:coulomb}
\ee
and gives an order $g^3$ contribution to the effective potential. In particular for $q=0$, this contribution is proportional to $m_D^3$.

  For the diagrams in fig.(\ref{fig:Df}) and fig.(\ref{fig:Dp}) to be finite we have to check that the selfenergy at small momenta behaves like ${\vec p}^2$, and this is indeed the case.

For the infrared finite result to be Z(N) invariant we should keep 
  contributions from all Matsubara frequencies in 
 the selfenergy. After doing so the selfenergy reads:
\be
\Pi_{c,c'}^{(k)}(q)=4g^2f^{c,a,b}\widehat B_2(C_a)f^{c',a,b}
\label{eq:PiB2}
\ee
\noindent  where $\widehat B_2(q)$ is related to the Bernoulli
polynomial:
\be
\widehat B_2(q)={T^2\over 2}({1\over 6}-|q|+q^2)\equiv {T^2\over 2}B_2(q)
\ee
This Bernoulli  polynomial is periodic mod 1, because we summed over all Matsubara frequencies in the selfenergy.
The formula is only true for external legs $c$ and $c'$ being neutral, i.e.
not feeling the background field. In that case $C_a=-C_b$. It is proven in  appendix B. Note that the dependence on the channel k comes in only through 
the argument $C_a$. 

The argument of $\widehat B_2(C_a)$ is 0 if the index a is that of a diagonal 
basis element, or $C_i-C_j$ if $a=(ij)$. The group structure constants are
defined as:
$$if^{a,b,c}=2Tr[\lambda^a,\lambda^b]\lambda^c$$ 

Eq.(\ref{eq:PiB2}) is diagonalized  for any tunneling path in  appendix C.

\begin{figure}
\begin{center}
\includegraphics{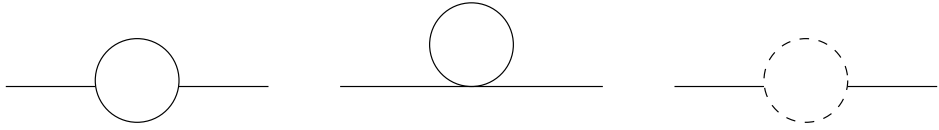}
\caption{The (00) component of the gluon selfenergy of  with fixed  colours a and b on the internal lines as in eq.(\ref{eq:PiB2}). See Appendix C for evaluation.}
\label{fig:self}
\end{center}
\end{figure}

Here we give the results for the eigenvalues $\Pi^{(k)}_e(q)$.
Obviously all eigenvalues have the Debye mass at $q=0$ and $q=1$ in common.
The selfenergy matrix contains in every channel $k$ the $N-1$ by $N-1$ matrix of diagonal
gluons. Furthermore the indices $c$ and $c'$ can take any value $ij$ for which the background $C_{ij}=0$ in the incoming propagator. Thus the  dimensionality of the matrix depends on k through the number of 
zero eigenvalues of the adjoint representation of $Y_k$: $N^2-1-2k(N-k)$.

In terms of $\widetilde B_2(q)\equiv B_2(q)-{1\over 6}$ and dropping the overall
scale factor $m_D^2$ one finds \linebreak
$(k-1)(k+1))$ eigenvalues
\be
(1+6{N-k \over N})\widetilde B_2(q)
\label{eq:ev1}
\ee
(N-k-1)(N-k+1) eigenvalues
\be
1+6{k \over N} \widetilde B_2(q)
\label{eq:ev2}
\ee
and one eigenvalue
\be
1+6\widetilde B_2(q)
\label{eq:ev3} 
\ee

\subsection{Infrared behaviour  of the Polyakov loop}\label{subsec:polyakovren}

The graph in fig.(\ref{fig:Dp}) shows the renormalization of the Polyakov loop
to $O(g^3)$.  The calculation parallels that in ref.~\cite{bhatta},~\cite{korthalsaltes}, but now with the resummed gluon propagator from eq. (\ref{eq:coulomb}), and the diagonalized version of eq.(\ref{eq:PiB2}) as described above.
The effect on the eigenvalues of the Polyakov loop matrix is given by:
\be
{\delta C_i-\delta C_j\over{2\pi}}={g^2N\over{(4\pi)^2}}(3-\xi)B_1({C_i-C_j\over{2\pi}})+ {g^2N\over{(8\pi)}}{m_D\over T}\Big(1+P_3\Big).
\label{eq:polyakovrenorm cube}
\ee

The second term is due to heavy modes of order $T$ and contains the gauge
dependence needed to cancel the gauge dependence in the  graphs of
fig. (\ref{fig:12loop}). The cubic term \footnote{Remember $m_D/T=({g^2N\over
    3})^{1/2}$.} contains as first term the one discussed by Jengo and
Gava~\cite{gava} without background field. The second term $P_3$ is induced by
the background. The precise dependence on the background we do know but it is
irrelevant for the purpose of this paper as the following argument shows.

The cubic term is inserted into the lowest order result eq. (\ref{eq:onelooppot}) and gives a cubic contribution to the tension: $\sum_{ij}\delta C_{ij} B_3(C_{ij})$. This cubic contribution, unlike the quadratic one, is zero! The reason is that the background contribution is only present in those combinations (ij) with $C_{ij}=0$, and $B_3(0)=0$. For the background independent correction
the sum $\sum_{ij}B_3(C_{ij})$ in the insertion cancels out on the path $qY_k$.
This is because $k(N-k)$ terms multiply $B_3(q)$ and an equal number $B_3(-q)$.  Since $B_3$ is an odd function~\cite{bhatta} the result cancels out.

\subsection{Infrared behaviour of the kinetic term to two loop order}\label{sec:kin2loop}

We still have to understand wether the two loop kinetic term can contribute
to cubic order. 
This term, nominally of $O(g^4)$, will contain contributions of order
$g^3$. The question is wether they are present for all values of $q$
including $q=O(1)$.
 So we will insist on at least one propagator with a background induced 
mass of order $T$, i.e. $q=O(1)$. From inspection of the diagrams
that contain vertex corrections one sees that this entails at least three
propagators with the mass of order T. Such diagrams are infrared finite.
The diagram with a selfenergy insertion has the same infrared properties 
as the one without insertion so will be infrared finite ~\footnote{or will contribute $g^3\log({1\over g})$ to $V$ if $q=O(g^2)$ so $O(g^7\log(1/g)$ to the tension}. 
In conclusion, no  two loop kinetic term contributes $O(g^3)$ to the tension.

\section{The $O(g^3)$ contribution to the effective potential}\label{sec:g3}

We have now eliminated the infrared divergencies from the potential up and including three loop order.
By resumming the Z(N) invariant selfenergy  $\Pi^{(k)}(q)$ this procedure
respects the Z(N) invariance of the potential.

\begin{figure}
\begin{center}
\includegraphics{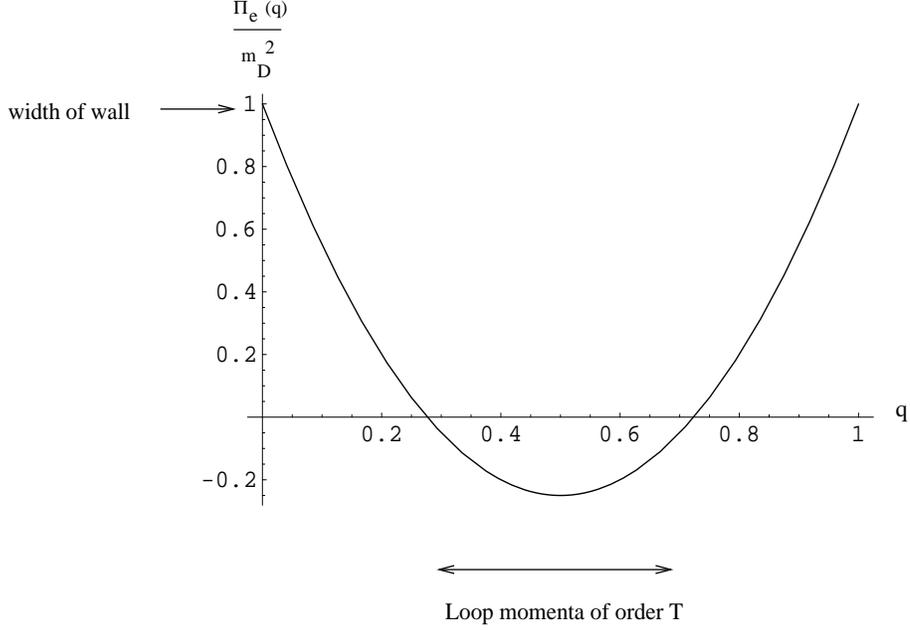}
\caption{The selfenergy for $r={5\over 6}$. Note that the selfenergy is maximal for $q=0$ and $q=1$ and equals there the (normalized) Debye mass.}
\label{fig:Pi}
\end{center}
\end{figure}

Although we have  managed to get rid of the divergencies and keeping Z(N) invariance, there is a price to pay!
Now the  selfenergy is negative  in a certain window of q values of order 1, see fig.(\ref{fig:Pi}).  So this window comes from integrating over hard momenta inside the selfenergy loop. The problem has not to do with the infrared
scales. It is a generic problem occurring in  quantum corrections to 
tunneling through a barrier~\cite{dashen}.  However,  for a window of loop strengths ${1\over 3}<{k\over N}<{2\over 3}$ the mass stays positive over the whole range of q values.  In particular in the large N limit we will be able to extract the potential without this problem.

From the result for the masses of the Coulomb propagators in the previous section we find the contribution to the free energy. It will be a sum of terms 
of the type
\be
g^3V_3(q)=\sum_e T\int {d\vec l\over{(2\pi)}^3}\log({\vec l}^2+\Pi_e(q))
\label{eq:contribution}
\ee
over all eigenvalues.

We write for the ratio ${k\over N}=r$ and introduce the function:

\be
f(r,q)=(1+6rq(q-1))^{{3\over 2}}-1
\label{eq:f}
\ee
Note that this function is non-positive.
The cubic potential, eq.(\ref{eq:contribution}),  becomes in terms of this function:
\be
g^3V_3=-{1\over{12\pi}}Tm_D^3N^2\big(f(r,q)(1-r)^2+f(1-r,q)r^2-{1\over{N^2}}\big(f(r,q)+f(1-r,q)-f(1,q)\big)\big)
\label{eq:finalfree}
\ee

All terms proportional to ${1\over N}$ cancelled out.
The leading terms, proportional to $N^2$, are well defined within the window
${1\over 3}\le r\le {2\over 3}$ as seen from eq.(\ref{eq:f}).

\section{The tension to cubic order}\label{sec:cubic}

The cubic term in the tension is obtained from the minimization. The relevant expression was given in eq.(\ref{eq:tensiongen}):
\be
\rho_k={2\over g}\int^1_0 dq (K(q)V(q))^{1\over 2}
\ee
Combine  the classical part of the kinetic term and the cubic 
part in $V$, eq.(\ref{eq:finalfree}) and integrate over q.

The tension becomes then to cubic order:
\be
\rho_r(T)=\rho_r^{(1)}\big(1- 1.0341..(\alpha_s N)+({3\over{\pi^3}})^{{1\over 2}}I(r)(\alpha_sN)^{{3\over 2}}+O(\alpha_s^2)\big)
\label{eq:finalresult}
\ee

\begin{figure}
\begin{center}
\includegraphics{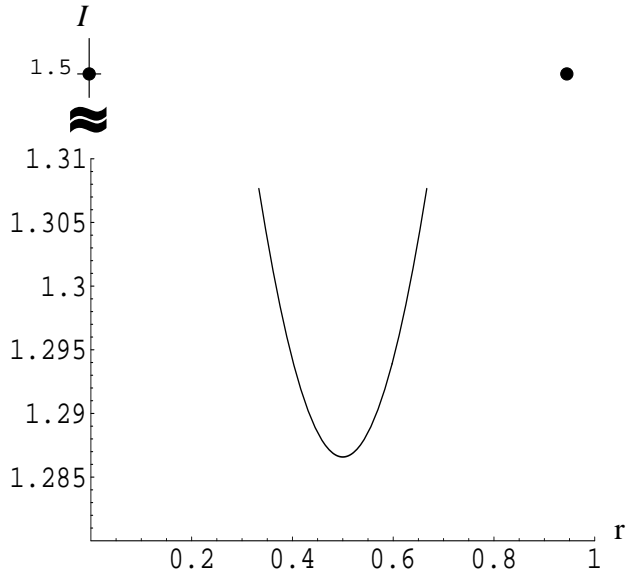}
\caption{The function I(r) in the result for the tension, eq.(\ref{eq:finalresult})}
\label{fig:ratio}
\end{center}
\end{figure}

The function 
\be
I(r)\equiv \int dq{(f(r,q)(1-r)^2+f(1-r,q)r^2)\over {6q(q-1)r(1-r)}}
\label{eq:Ifunction}
\ee
\noindent is shown in fig.(\ref{fig:ratio}). It shows a small deviation from k-scaling, as it drops from the value 1.5 at r=0 to $\sim 1.3$ at r=0.5. As in the pressure~\cite{recentkaj} it contributes with a sign opposite
to the $O(g^2)$ term.
 
The attraction between loops becomes stronger due to the convexity of $I(r)$.
To see this, take the ratio of the tension of the k-loop and compare it to
the k times the tension of the elementary loop with k=1:
\be
{\rho_{k\over N}(T)\over{k\rho_{1\over N}(T)}}=(1-r)\big(1+({3\over{\pi^3}})^{{1\over 2}}(I(r)-I(0))(\alpha_sN)^{{3\over 2}}+O(\alpha^2\big)+O({1\over{N^2}})
\ee
The point is that the cubic correction has a negative coefficient, so that the ratio is smaller due to the presence of this correction.

\section{Discussion}\label{sec:discussion}

We have computed the cubic correction to the tension of the 't Hooft loop.
At first sight the occurrence of such a term seems counter-intuitive. One would expect the profile to provide an infrared cut-off on the order of $|q|=O(T)$
well inside the wall. Only where the profile merges with the plasma ground state one expects the infrared divergencies to occur, as we discussed in section(\ref{sec:semiclass}) for the next to leading correction to the Debye mass.

This expectation is refuted by the presence of neutral gluonfields. They do
propagate as massless fields in a constant background. But in the wall they can
produce pairs of charged fields and interact therefore with the wall.
The latter fact suggests that one should put the profile into the selfenergy,
and then determine the free energy of the wall by computing the determinant of the operator. This method works not only for large N but for all SU(N) gauge theories. The mathematics of this method was used by Dashen, Hasslacher and Neveu~\cite{dashen}. It can be worked out analytically and will be the subject of a future publication~\cite{korthalsgiovanna}.

 Concerning the  magnetic components of the neutral gluons: do
they get a self mass? As we show in the   Appendix  the magnetic selfenergy vanishes also in the presence of the wall. So a neutral magnetic gluon stays neutral, even in the presence of the wall. This means that their effects are only calculable by non-perturbative means. Or, what amounts to the same, to obtain a mass, one would need to put in a spatial Wilson loop.

 Still, it may be that all logarithmic effects of the type we encountered for the Debye mass can be captured by the mass induced into the off-diagonal magnetic glue. In other words the definition of the free energy in the bulk as a limit
of our effective action where the Polyakov loop goes to its bulk value is not only manifestly gauge invariant but may also provide a well defined procedure to
compute those logarithms.

A last remark: our calculation applies to the domain walls discussed in ref.~\cite{korthalslaine}. In five dimension the effects are $g^4log(1/g)$ instead of $g^3$.
 
\section*{Acknowledgments}
P.G. thanks the MENESR for financial support. CPKA the CERN theory division for its hospitality and useful discussions with Mikko Laine and Philippe de Forcrand.  Discussions at SEWM 2002 in Heidelberg were fruitful, especially with  Rob Pisarski and Larry Yaffe.
  
\section*{Appendix A: k-scaling of the two loop contribution}

In this appendix we prove the factor $k(N-k)$ in front of eq.(\ref{eq:tensionordertwo}). The diagrams in fig.(\ref{fig:12loop}b) are all reduced by energy momentum conservation
to the form:
\be
\sum_{a,b,t}f^{a,b,t}f^{a,b,t}\sum\int {1\over{l^2_a}}\sum\int {1\over{l^2_b}}=\sum_{a,b,t}f^{a,b,t}f^{a,b,t}\widehat B_2(C_a)\widehat B_2(C_b)
\label{eq:groupsum}
\ee
in an evident notation ($C_d=0,C_{ij}=C_i-C_j$).
Split the sum over the colour factor $t$ into  $t=d$, $d$ diagonal and  $t=(i\neq j)$.

\noindent 1)In the first sum $a=ij$, and $b=ji$, not diagonal. Then the completeness relation:
\be
\sum_{d=2}^Nf^{ij,ji,d}f^{d,kl,lk}={1\over 2}(\delta_{ik}+\delta_{jl}-
\delta_{il}-\delta_{jk})
\ee

\noindent says the first sum equals 
$$\sum_{ij,ji,d}f^{ij,ji,d}f^{ij,ji,d}\widehat B_2(C_{ij})\hat B_2(C_{ji})
=2k(N-k){(\widehat B_2(q))}^2$$ because on the path $Y_k(q)$ $C_{ij}=\pm q$ for
 $2k(N-k)$
combinations and because  the sign of the argument $C_{ij}$ is
 immaterial in the even function $\hat B_2$.\\

Now the sum of $t$ over $ij$.  There are two different contributions.\\

\noindent 2a)First there is either $a=d$ or $b=d$.
Both give the same contribution $2k(N-k)\hat B_2(0)\hat B_2(q)$. \\

\noindent 2b)Then there is $a=jl$, and $b=li$.
In case the indices i and j are in different sectors of $Y_k$
either $l$ equals one of the $N-k$ indices $i$ or one of the $k$ indices $j$. 
In either
case the summand equals $\hat B_2(0)\hat B_2(q)$. And the coefficient in front is
$2k(N-k)(N-2)$, the factor $N-2$ coming from the $N-2$ ways we can put $l$.

In case the the indices $i$ and $j$ are in the same sector we get
$$\big(k(k-1)(N-k)+(N-k)(N-k-1)k\big){\widehat B_2(q)}^2$$
Adding up the contributions from 1), 2a and b) we get finally for the q dependent part:
\be
\sum_{a,b,t}{|f^{a,b,t}|}^2\widehat B_2(C_a)\widehat B_2(C_b)=Nk(N-k)\big(\widehat B_2(q)^2+2\widehat B_2(0)\widehat B_2(q)\big)
\ee

The contribution from the diagram in fig.(\ref{fig:12loop}c) equals:
\be
-4g^2Nk(N-k)\widehat B_1(q)\widehat B_3(q)  
\ee
where the first factor is the renormalization of the Polyakov loop
and the second the derivative of the one loop potential.

Added, they give the result in the text, eq.(\ref{eq:twolooppot}).

\section*{Appendix B: the selfenergy of the gluon in nontrivial background}

We want to prove eq.(\ref{eq:PiB2}) and show that the magnetic part of the selfenergy vanishes at zero momentum in the presence of the wall.
Look at fig.(\ref{fig:self}). The question is, how do we identify the summand for fixed
indices a and b in eq.(\ref{eq:PiB2})? For the first two graphs in fig.(\ref{fig:self}) this is rather obvious: the two internal lines carry the index a and b and will therefore have propagators ${1\over{l_a^2}}\equiv {1\over{(2\pi nT-C_a)^2+{\vec l}^2}}$ with $C_a=-C_b$. For the tadpole graph we symmetrize the sole propagator in a and b:
\be
f^{c,a,b}f^{c',a,b}\sum_n{1\over {l_a^2}}=f^{c,a,b}f^{c',a,b}\sum_n{1\over 2}\big({1\over {l_a^2}}+{1\over {l_b^2}}\big)
\label{eq:symmetrize}
\ee
In this formula the sum is over Matsubara frequencies only, not over colour indices. So the equality results from changing in the second term on the r.h.s.
the summation from n into -n and using $C_a=-C_b$.
Then one finds:
\be
\widehat B_2(C_a)=-T\sum_n\int{d\vec l\over{(2\pi)^3}}(4{l_a^0}^2-2l_a^2){1\over{l_a^2}^2}
\ee

Use 
\be
T\sum\int {d\vec l\over{(2\pi)^d}}{{\vec l}^2\over{(l_a^2)^2}}={d\over 2}T\sum\int {d\vec l\over{(2\pi)^d}}{1\over{l_a^2}}
\label{eq:identity}
\ee
\noindent to find indeed:
\be
\widehat B_2(C)={T^2\over 2}B_2(C)
\ee

The formula for the selfenergy in tensor form reads:
\be
\Pi_{\mu\nu}(C)=-T\sum_n\int{d\vec l\over{(2\pi)^3}}(4l_a^{\mu}l_a^{\nu}-2l_a^2\delta_{\mu\nu}){1\over{l_a^2}^2}
\ee
Contract the purely spatial part with $\delta_{ij}$ and use eq.(\ref{eq:identity}) to find it is identically zero, for any value of $C$ and $d$. This means that the wall induces only a Z(N) invariant Debye mass on the Coulomb fields. 

\section*{Appendix C: eigenvalues of  the selfenergy matrix}

We will diagonalize the selfenergy matrix, eq.(\ref{eq:PiB2}) in the main text : 
\be
\Pi_{c,c'}^{(k)}(q)=4g^2f^{c,a,b}\widehat B_2(C_a)f^{c',a,b}
\ee

This matrix has no overlap between $c=ij$ and $c'=d$ sectors as one can easily
from the routing the colour indices in the self energy diagrams.
So the matrix is divided into two sectors.

 The first sector is determined by the  subset  $c=ij$ and $c'=mn$ defined by those phases $C_{ij}$ being identically zero on the tunneling path $Y_k(q)$.
Recall that $$C_{ij}={2\pi\over N}((Y_k(q))_{ii}- (Y_k(q))_{jj})$$
With the definition $Y_k(q)=qY_k$ we find therefore $C_{ij}=0 \mbox{ or} \pm2\pi q$. Here we are interested in $ij$ such that $C_{ij}=0$. There are $O(N^2)$ 
such pairs.

The second one is the one
where the two propagators in the diagram are colour diagonal. It will be a $N-1$ by $N-1$ matrix and will differ from one tunneling path to the other. The second sector involves only
$O(N)$ degrees of freedom and will be subdominant.

\null

{\bf{ Eigenvalues in the off-diagonal colour sector}}\\


The one loop self energy $\Pi^{(k)}_{ij,mn}(q)$ is diagonal in $ij$ and $mn$. So its value
can be written, from eq.(\ref{eq:PiB2}), as:
\be
\Pi^{(k)}_{ij}(q)=2\sum_{h} g^2|f^{ij,jh,hi}|^2T^2(B_2(0)+\tilde B_2(C_{hj}))
\label{eq:offdiag}
\ee
Obviously after summing over $h$ the Debye mass $NB_2(0)$ should appear at $C_{hj}=0$ or $1$. 
 
In the channel $k$ one has two cases:\\
\null
\noindent i)The index pair (ij) is from the first $N-k$ entries in the charge matrix $Y_k$. Then the running index h must be in the last k entries of $Y_k$ , to give a nonzero contribution $\tilde B_2(C_{ih}=\tilde B_2(q)$. So $B_2(0)$ in eq.(\ref{eq:offdiag}) is multiplied by N, the N possible values of the index h. But $\tilde B_2(C_{ih})$ is non zero only for the k values of the index h. So the result is:  
\be
\Pi^{(k)}_{ij}(q)=2g^2T^2(NB_2(0)+k\tilde B_2(q))=m_D^2(1+6k\tilde B_2(q))
\ee 
\noindent for any of the $(N-k)(N-k-1)$ pairs $ij$.
\null\\
\noindent ii)The index pair (ij) is from the last k entries in $Y_k$. Then the index h 
must take on one of $N-k$ values in the matrix $Y_k$ and the result becomes:
\be
\Pi^{(k)}_{ij}(q)=2g^2T^2(NB_2(0)+(N-k)\tilde B_2(q))=m_D^2(1+6(N-k)\tilde B_2(q))
\ee
\noindent for any of the $k(k-1)$ index pairs $ij$.

\null

{\bf{ Eigenvalues in the diagonal colour sector}}\\   

\null

First a few general considerations.
 Of course  any tunneling path differing from $Y_k(q)$ by a permutation $P$ of the diagonal elements of $Y_k(q)$ is physically the same: the potential should be identical.
It is not hard to show that any such permutation amounts to an orthogonal
transformation $O$ acting on the orthogonal basis of the $\lambda_d$.
So in particular the structure constants $f^{d,ij,ji}$ transform like:
\be
f^{d,P(ij),P(ji)}=O_{d,d'}f^{d',ij,ji}
\ee
Now our selfenergy matrix reads:
\be
\Pi^{(k)}_{d,d'}(q)=4g^2f^{d,ij,ji}\widehat B_2(C_{ij})f^{d',ij,ji}
\ee
where we sum over the $ij$ indices.

 It is then clear that the selfenergy transforms into $O\Pi^{(k)}O^{-1}$, when computed from $B_2(C_{P(ij)})$ instead of $B_2(C_{ij})$. So its eigenvalues stay the same. And only the eigenvalues appear in the potential. 
Another property is that the eigenvalues are the same for charge conjugate paths $Y_k(q)$ and $Y_{N-k}(q)$, as  expected. This comes about because a permutation changes one into
the other up to a minus sign. This sign shows up in the argument of $B_2(C_{ij})$
but is immaterial because $B_2$ is even.
Like before it is useful to split $B_2$ into its value at $q=0$ and the rest:
\be
B_2(q)=B_2(0)+\tilde B_2(q)
\ee
The first term contributes for every pair $ij$ in the sum that constitutes the mass matrix. Hence it comes in as proportional to the unit matrix with proportionality constant ${m_D^2\over 2}$.
The remaining term generates the non-trivial part of the selfenergy matrix:
\be
2\sum_{ij}f^{d,ij,ji}g^2T^2\tilde B_2(q)f^{d',ij,ji}
\ee

In contrast to the q-independent part this part
of the mass matrix is in general non-diagonal. For $k=1$ it is diagonal,
for $k=2$ it has off diagonal elements between $d=N$ and $d=N-1$, for $k=3$ 
between $d=N, N-1\hbox { and }N-2$. 
After diagonalization the non-trivial mass matrix has the  eigenvalues (dropping the common factor $6m_D^2\tilde B_2(q)$:
\be
k
\ee

with multiplicity $N-k-1$.

Then it has $k-1$ eigenvalues
\be
(N-k)
\ee
\noindent and one eigenvalue
\be
N
\ee

As dictated by charge conjugation and periodicity modulo N the above set of eigenvalues does not change when we exchange k and N-k.

A useful check on this result comes from taking  the trace of the non-trivial selfenergy matrix, using $\sum_d|f^{d,ij,ji}|^2=1$.
  So this trace counts the number of nonzero eigenvalues in the adjoint representation of $Y_{k}$, which is $2k(N-k)$. So is the sum of the eigenvalues we found by explicit diagonalization of the selfenergy matrix above.

Taking the results for the eigenvalues and multiplicities together one arrives 
at the set of eigenvalues and multiplicities in section(\ref{sec:threeloop}).

\end{document}